# Universal Route for the Emergence of Exceptional Points in PT-Symmetric Metamaterials with Unfolding Spectral Symmetries

Yanghao Fang[1], Tsampikos Kottos[2,†], and Ramathasan Thevamaran[1,3,*]

[1]*Department of Materials Science and Engineering, University of Wisconsin-Madison, Madison, Wisconsin 53706, USA*
[2]*Wave Transport in Complex Systems Lab, Department of Physics, Wesleyan University, Middletown, Connecticut 06459, USA*
[3]*Department of Engineering Physics, University of Wisconsin-Madison, Madison, Wisconsin 53706, USA*

[†]tkottos@wesleyan.edu
[*]thevamaran@wisc.edu

## Abstract

We introduce a class of Parity-Time symmetric elastodynamic metamaterials (Ed-MetaMater) whose Hermitian counterpart exhibits a frequency spectrum with unfolding (fractal) symmetries. Our study reveals a scale-free formation of exceptional points (EP) whose density is dictated by the fractal dimension of their Hermitian spectra. Demonstrated in a quasi-periodic Aubry-Harper, a geometric H-tree-fractal, and an aperiodic Fibonacci Ed-MetaMater, the universal route for EP-formation is established via a coupled mode theory model with controllable fractal spectrum. This universality will enable the rational design of novel Ed-MetaMater for hypersensitive sensing and elastic wave control.

## Introduction

Distinct from common geometric symmetries of phononic crystals and metamaterials, the Parity-Time (PT)-symmetric materials [1–4] utilize hidden symmetries that are encoded in the governing dynamical equations and are consequences of judicious spatially-distributed attenuation and amplification mechanisms. The PT-symmetric systems have been shown to exhibit novel transport phenomena in various application domains such as optics [5–9], microwaves and radiofrequency waves [10–13], and acoustics [4,14–19]. Unidirectional invisibility [16,17], shadow-free sensing [15], asymmetric switching [4,11], and non-reciprocal transport [20,21] are some of those exotic wave phenomena that have been demonstrated both theoretically and experimentally. On the other hand, very few works have been focused on the implementation of PT-symmetry in the realm of elastodynamics concerning the elastic wave dynamics in solids [22,23].

A PT-symmetric elastodynamic metamaterial (Ed-MetaMater) has recently been realized by embedding a gain and a lossy mechanical resonators in an elastic medium that facilitates coupling between them [22]. When the intensity of the equal gain/loss and/or the elastic coupling strength between the two resonators of such an Ed-MetaMater are varied, a branch-point singularity forms where the eigenmodes and eigenvalues of the system coalesce. Such degeneracy is known as an exceptional point (EP) and it is the most intriguing feature of PT-symmetric systems: It signifies a transition from a parameter domain where the eigenfrequencies are real and the corresponding eigenmodes of the system respect the PT-symmetry (exact PT-symmetric phase) to a domain where the eigenfrequencies are complex conjugate pairs and the normal modes violate the PT-symmetry (broken PT-symmetric phase) [24,25]. In the vicinity of an EP, the degenerate eigenfrequencies can be expanded in a fractional (Puiseux-Newton) power series whose importance in sensing applications has been recognized only recently [26–29]. The EP degeneracies have so far been implemented using coupled resonators in zero (e.g. pair of coupled resonators) or one-dimensional geometries (e.g. one-dimensional arrays of coupled resonators). Developing methods that allow the implementation of EPs in more complex geometries will provide exciting opportunities to engineer mechanical wave dynamics.

The quasi-periodic, aperiodic, and geometric fractal architectures offer new ways of engineering metamaterials across multiple length scales and response time scales because of their intriguing frequency spectrum demonstrating unfolding (fractal) symmetries [14,30–34]. The embodiment of fractality in PT-symmetric metamaterials offers the potential to create numerous EPs in a scale-free fashion similar to the scale-free nature of their fractal frequency spectra. Here, we introduce PT-symmetric Ed-MetaMaters with fractal frequency spectra and establish a universal route for the emergence of EPs in those metamaterials. We describe the mechanisms of the emerging EPs induced by the elastodynamic interactions in three classes of PT-symmetric Ed-MetaMater with fractal spectrum—a quasi-periodic (incommensurate) Ed-MetaMater inspired by the Aubry-Harper model [35–41] (Fig.1(a)), a geometric fractal Ed-MetaMater made of H-shaped motifs [42] (Fig.1(b)), and an aperiodic Ed-MetaMater that follows Fibonacci substitutional rule (Fig.1(c)) [43,44]. The universal scale-free nature of the EP formation and its connection to the fractal dimension of the frequency spectrum of the underlying Hermitian Ed-MetaMater are established using a coupled-mode-theory (CMT) modeling (Fig.1(d)).

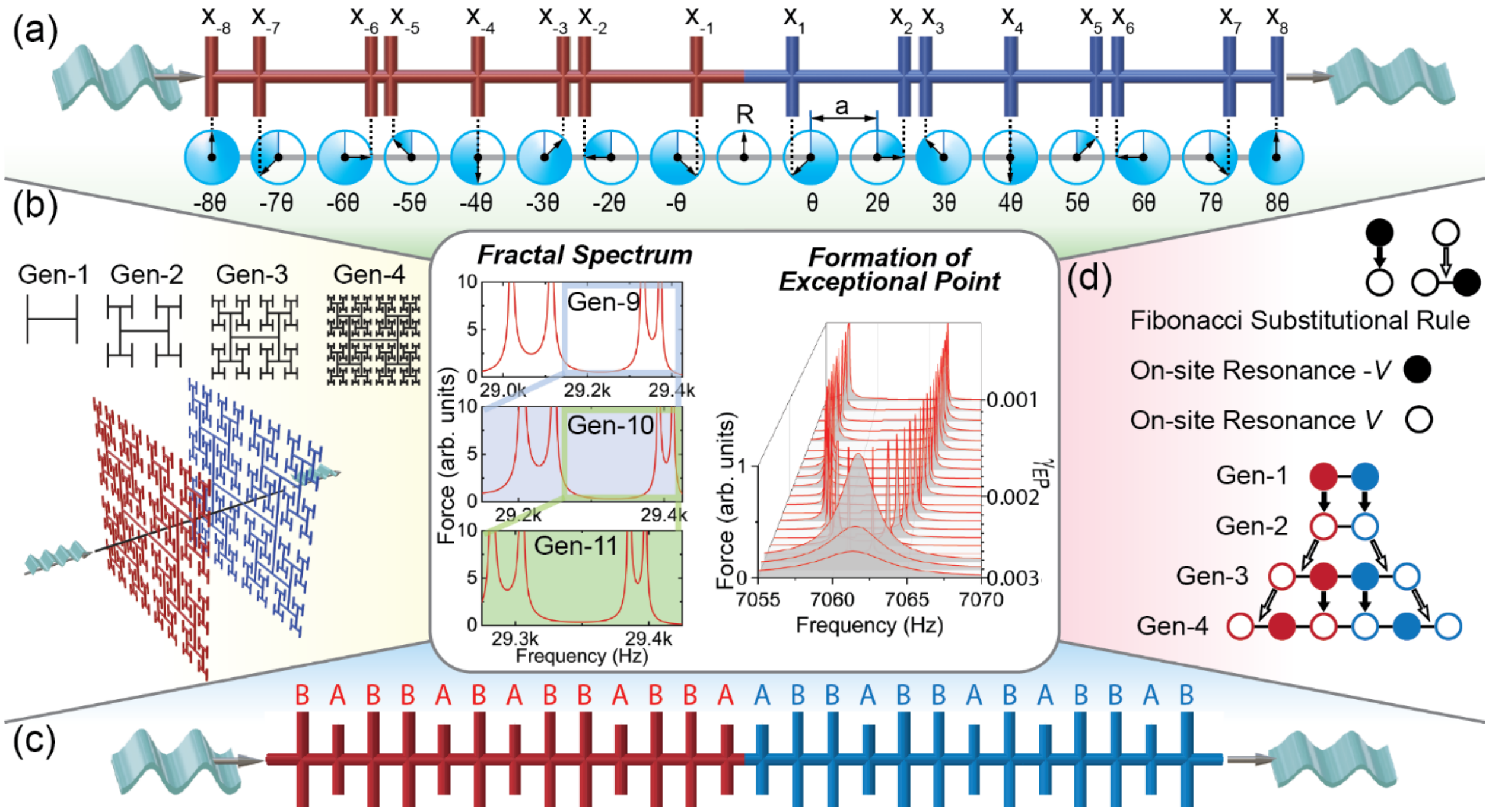

FIG.1. The Scale-free emergence of EPs in PT-symmetric Ed-MetaMaters displaying self-similar fractal spectra (red/blue colors represent energy gain/loss). (Central image) Self-similarity of fractal spectrum as generation is increased and a representative EP formation. (a) A Generation 7 (Gen-7) Aubry-Harper Ed-MetaMater. (b) A Gen-4 H-tree-fractal Ed-MetaMater and first four generations of its planar component. (c) A Gen-7 Fibonacci Ed-MetaMater. (d) The first four generations of the CMT-based mathematical model with on-site resonances following Fibonacci substitutional rule.

**The PT-symmetric quasi-periodic Aubry-Harper Ed-MetaMater**

The finite element model of the quasi-periodic Aubry-Harper Ed-MetaMater is generated by several vertical beams (length: 6 cm; radius: 0.5 cm) coupled by a horizontal rod with their positions given by $x_s = sa + R\sin(s\theta)$, where $s \in \mathbb{Z}$ is the rod index, $a$ ($5\ cm$) is the distance between the centers of neighboring circles, and $R$ ($2\ cm < a/2$) is the radius of the circle (Fig.1(a)). When the projection parameter $\theta \in (0,1)$ is an irrational number, the period of the impedance profile of the structure is incommensurate with the lattice period. To this end, we use the ratio of two adjacent numbers in a Fibonacci sequence for $\theta = p/q$, so that the impedance profile of the structure becomes commensurate with the lattice of rods with period $q$, defining the generation of this Ed-MetaMater (e.g., 7$^{\text{th}}$ generation (Gen-7) corresponds to $\theta = 5/8$, which is composed of the 6$^{\text{th}}$ and 7$^{\text{th}}$ numbers in Fibonacci sequence: 0, 1, 1, 2, 3, 5, 8, 13, ….). The incommensurate limit associated with a truly quasi-periodic structure is investigated via a scaling procedure and it is reached when $q \to \infty$. A Gen-7 PT-symmetric Aubry-Harper Ed-MetaMater consists of two parity (P)-symmetric components shown in red/blue in Fig.1(a), indicating spatially localized balanced energy gain/loss ($-/+\gamma$).

The P-symmetric Aubry-Harper Ed-MetaMater is harmonically excited (0-25 kHz) using a prescribed axial displacement at the left-end of the horizontal coupling rod and the corresponding sinusoidal axial reaction force at its fixed-right-end is measured (details and Fig.S1 in [45]), simulating a steady-state elastic wave dynamics. The frequency spectrum of its response shows numerous resonance modes that emerge in a scale-free nature as the generation of the Ed-MetaMater is increased (central figure in Fig.1). The fractal dimension $D$ of the frequency spectra of the P-symmetric ($\gamma = 0$) Aubry-Harper Ed-MetaMater is calculated using the correlation-dimension method [46] and found to be $D = 0.83 \pm 0.02$ (Fig.S2(a) [45]) (a standard box-counting method results in the same $D$—albeit the correlation-dimension method converges faster to the value of $D$). We also find that the $D$ of the P-symmetric Ed-MetaMater is the same as that of the corresponding Ed-MetaMater without being coupled to its mirror image. The $D$ remains the same even in the PT-symmetric case—albeit in this case, it refers only to the real part of the frequencies. This robustness of the $D$ of the real part of frequency spectrum against P or PT-symmetries was checked for all the systems we studied.

The PT-symmetry is created by introducing at each P-symmetric part of the Ed-MetaMater equal amount of energy gain and loss ($-/+\gamma$), characterized by an amplification/attenuation rate. This gain/loss is modeled in finite element model by a structural anti-damping/damping coefficient ($\gamma = 0.001 \sim 1$). When the $\gamma$ is increased, several pairs of modes interact and coalesce to form a cascade of EPs at different critical values $\{\gamma_{EP}^{(n)}\}$, and a square-root behavior typical of order-two EPs can be observed near the EP (Fig.S2(b) [45]).

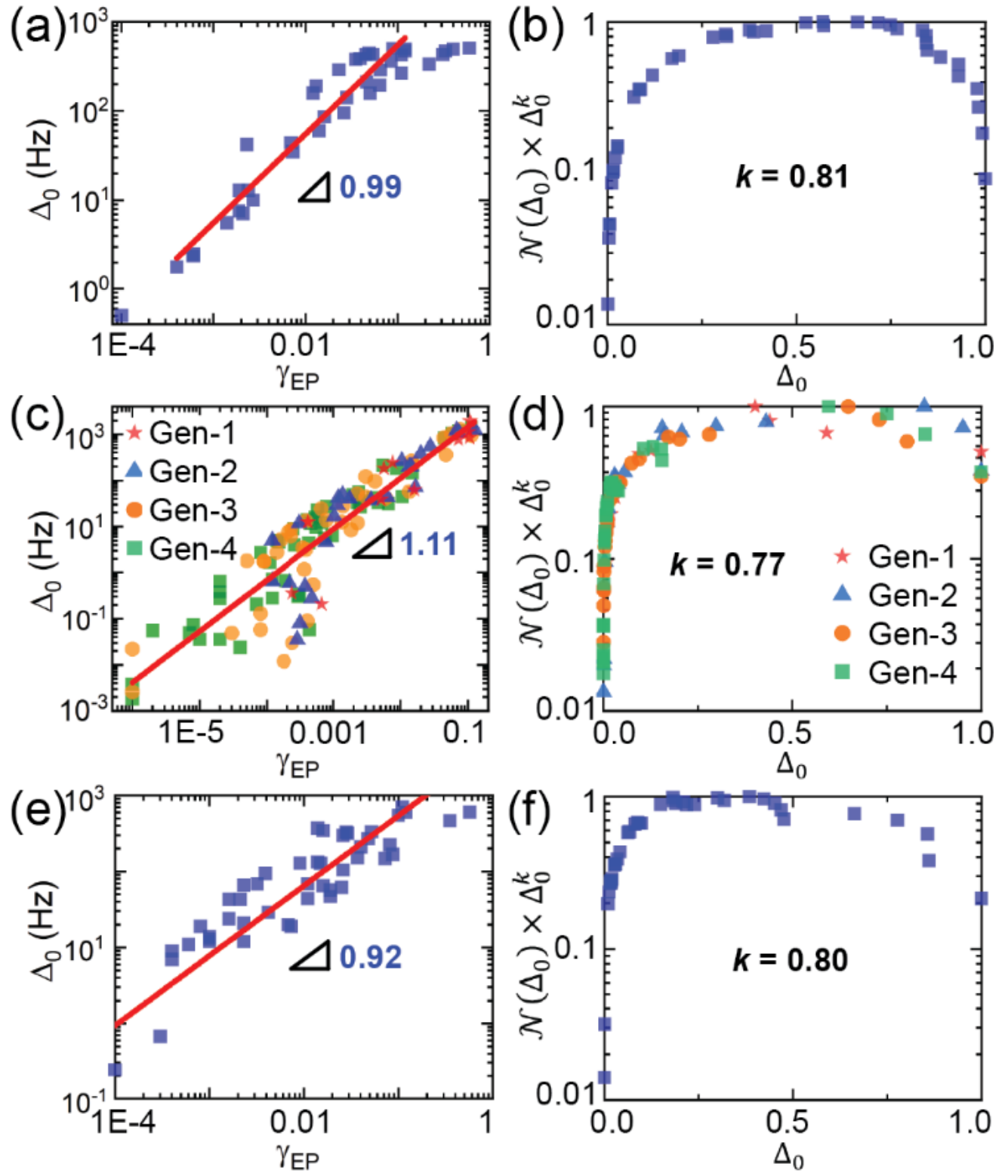

FIG.2. The EPs in PT-symmetric (a-b) Gen-10 Aubry-Harper, (c-d) H-tree fractal, and (e-f) Gen-9 Fibonacci Ed-MetaMater. (a,c,e) Linear relationship between $\Delta_0$ and $\gamma_{EP}$. (b,d,f) Integrated distribution reported as $\mathcal{N}(\Delta_0) \times \Delta_0^k$; axes normalized by their corresponding maxima.

To further estimate the non-Hermitian $\gamma$ that enforces an EP degeneracy for a specific pair of modes, we plot $\gamma_{EP}$ vs. the frequency difference between the corresponding mode pairs when $\gamma = 0$ ($\Delta_0 \equiv \Delta_{EP}|_{\gamma=0}$) for each EP found in the spectrum. All EPs ($< 25\ kHz$) in a Gen-10 PT-symmetric Aubry-Harper Ed-MetaMater are shown in Fig.2(a). Their linear relation demonstrates the intimate relation between the initial (i.e. when $\gamma = 0$) frequency split of these two interacting modes $\Delta_0$ and the critical gain/loss intensity $\gamma_{EP}$ which coalesce those modes to form an EP. In other words, the non-Hermitian perturbation strength $\gamma_{EP}$ that is needed for enforcing a degeneracy between an EP-pair must be of the same order as the frequency split of those modes in the P-symmetric Ed-MetaMater. Therefore, a statistical analysis of $\gamma_{EP}$ reduces to the statistical description of these $\Delta_0$, which are associated with the specific mode pairs that eventually form EPs. The latter is easier to evaluate numerically since it does not require a high-resolution parametric evaluation of the modes—as opposed to the precise determination of $\gamma_{EP}$.

We evaluate the probability density function (PDF) $\mathcal{P}(\Delta_0)$ of those EPs. For better statistical processing of these data, we refer to the integrated distribution $\mathcal{N}(\Delta_0) = \int_{\Delta_0}^{\infty} \mathcal{P}(x)dx$ whose derivative $\mathcal{P}(\Delta_0) = -d\mathcal{N}(\Delta_0)/d\Delta_0$ determines the PDF of the frequency split of the EP-pairs and therefore the PDF $\mathcal{P}(\gamma_{EP})$ of the gain/loss intensity that is necessary for inducing an EP degeneracy. We find that

$$\mathcal{N}(\Delta_0) = \int_{\Delta_0}^{\infty} \mathcal{P}(x)dx \sim \Delta_0^{-k}. \quad (1)$$

where the best fit parameter $k$ is found to be $k = 0.81 \approx D$. Thus the PDF for the gain/loss intensity scales as $\mathcal{P}(\gamma_{EP}) \sim \gamma_{EP}^{-(1+D)}$, which is represented by the flat spread in $\mathcal{N}(\Delta_0) \times \Delta_0^k$ vs. $\Delta_0$ (Fig.2(b)).

**The PT-symmetric H-tree geometric fractal Ed-MetaMater**

We investigate another class of Ed-MetaMater whose fractal spectrum is originating from a geometric fractality in configuration space [42]. This metamaterial is made of two identical planar components in H-motifs (Fig.1(b)). The first generation of this fractal contains two H-shaped structures (indicated in red and blue)—each made of 3 identical cylindrical beams (length: 11.6 cm, radius: 2.38 mm)—coupled by a passive (zero gain/loss) horizontal elastic rod (coupling length: 11.6 cm; length of exterior side ledges: 5.8 cm). Each subsequent generation adds H-motifs scaled down in length by a factor of 2 (constant diameter) to each tip of the prior H-structure (Fig.1(b)). The analysis of the correlation-dimension of the frequency spectrum indicates that its $D$ converges to 0.80 (Fig.S3 [45]).

As previously, the PT-symmetry is created by introducing equal gain/loss ($\gamma = 0.001 - 1$) to the left and right planar components (Fig.1(b)). The P-symmetric H-tree-fractal Ed-MetaMater is harmonically excited (0-50 kHz) similar to Aubry-Harper (Fig.S4 [45]). When $\gamma$ is increased, similar to Aubry-Harper, all four generations of the PT-symmetric H-tree-fractal Ed-MetaMater show the emergence of numerous EPs at $\left\{\gamma_{EP}^{(n)}\right\}$ which are proportional to the $\Delta_0$ associated with those specific EP-pairs (Fig.2(c)). We evaluated the $\mathcal{N}(\Delta_0)$ which allows estimating the PDF $\mathcal{P}(\gamma_{EP})$ for the critical gain/loss intensity $\gamma_{EP}$. Fig.2(d) shows the integrated distribution by the variable $\mathcal{N}(\Delta_0) \times \Delta_0^k$ as a function of $\Delta_0$. We find that for $k = 0.77 \approx D$, the data demonstrate a flat spread, leading to the conclusion that $\mathcal{P}(\gamma_{EP}) \sim \gamma_{EP}^{-(1+D)}$. This finding again demonstrates the intimate relation between the emerging EPs and the fractality of the metamaterial's spectrum.

**The PT-symmetric aperiodic Fibonacci Ed-MetaMater**

To further verify the universal nature of Eq.(1) we studied another class of Ed-MetaMater with unfolding spectral symmetries—an aperiodic system based on Fibonacci substitutional rule (Fig.1(c) and details in [45]). Equally spaced ($3\ cm$) vertical beams (lengths of A: $8\ cm$, B: $6\ cm$; radius: $0.5\ cm$) are coupled by a horizontal rod (radius: 0.5 cm; exterior ledges: $3\ cm$) and the system is harmonically (0-50 kHz) excited (Fig.S5 [45]). The correlation-dimension analysis indicates that the frequency spectrum of this system is characterized by $D = 0.80$ (Fig.S6 [45]). The relation between $\Delta_0$ and $\gamma_{EP}$ is found to be linear again (Fig.2(e)). The $\mathcal{N}(\Delta_0) \times \Delta_0^k$ vs. $\Delta_0$ demonstrates a flat spread with $k = 0.80 \approx D$ (Fig.2(f)), concluding that $\mathcal{P}(\gamma_{EP}) \sim \gamma_{EP}^{-(1+D)}$.

**A universal mathematical model for PT-symmetric fractal metamaterials**

The intimate relation between $\mathcal{P}(\gamma_{EP})$ and the spectral fractal dimension of an Ed-MetaMater at $\gamma = 0$ implies the existence of an underlying universal route for the creation of EPs in systems with fractal spectrum. To this end, we develop a CMT-based model that utilizes on-site resonant modes that follow an aperiodic Fibonacci substitutional rule (details in Fig.1(d) and [45]). The CMT Fibonacci model is described by the Hamiltonian:

$$H = \sum_n |n\rangle V_n \langle n| + \sum_n |n+1\rangle\langle n| + c.c. \quad (2)$$

where the coupled resonant frequencies $V_n$ take only two values $\pm V$ arranged in a Fibonacci sequence and $\{|n\rangle\}$ is the local mode basis. This system is known to have a Cantor-set spectrum with zero Lebesgue measure for all $V > 0$ [47]. Moreover, its spectral fractal dimension $D$ can be tuned by varying on-site resonance of the model, $V$. It turns out that the PDF of the nearest level spacing $s_n \equiv \omega_{n+1} - \omega_n$ of such family of systems follows a scale-free distribution whose power-law behavior is dictated by the fractality of the spectrum, i.e. $\mathcal{P}(s) \sim s^{-(1+D)}$ [48–50]. This power law is a signature of level clustering and it is distinct from the PDF $\mathcal{P}(s)$ of chaotic or integrable

systems [51,52]. We point out that the realization of this class of systems is not confined only to aperiodic systems like the Fibonacci chain model in Eq.(2), but also applicable to quasi-periodic systems with metal-insulator transition at some critical value of the on-site resonance (e.g. the Aubry-Harper model) [32,53,54], or wave systems with a chaotic classical limit as the kicked Harper model [52]. Therefore, our CMT model represents a typical example of a whole class of systems with fractal spectrum.

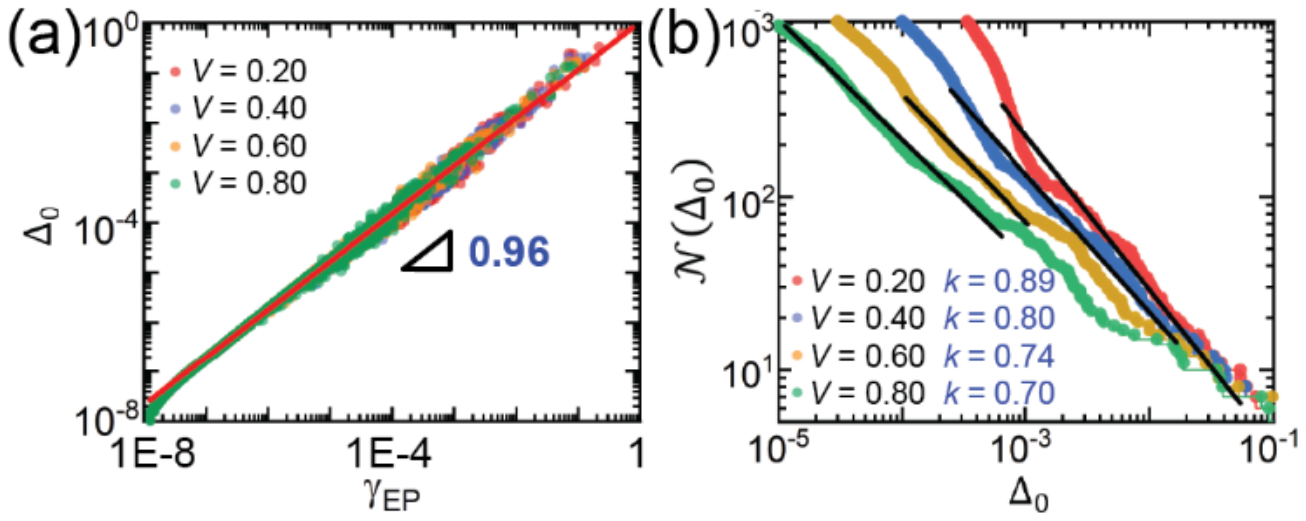

FIG.3. The EPs in the CMT-based mathematical model with Fibonacci on-site resonances $V$. (a) The linear relation between $\Delta_0$ and $\gamma_{EP}$. (b) Integrated distribution $\mathcal{N}(\Delta_0)$ as functions of $\Delta_0$ in Gen-17 Fibonacci CMT model.

The P-symmetric Fibonacci model is implemented by coupling the Hamiltonian of Eq.(2) with its mirror image. The corresponding effective CMT Hamiltonian takes the form:

$$H_P = \left(\sum_{n=-N}^{-1} |n\rangle V_n \langle n| + \sum_{n=-N}^{-2} |n+1\rangle\langle n| + cc\right) + \left(\sum_{n=1}^{N} |n\rangle \bar{V}_n \langle n| + \sum_{n=1}^{N-1} |n+1\rangle\langle n| + c.c.\right) + (|1\rangle\langle -1| + c.c.). \tag{3}$$

where $\{\bar{V}_n\}$ is the mirror-symmetric Fibonacci sequence of $\{V_n\}$ and $t$ describes the coupling between two Fibonacci chains. We found that the P-symmetric variant has a fractal frequency spectrum with the same $D$ as the one of the systems of Eq.(2). Finally, a PT-symmetric CMT model $H_{PT}$ is implemented by introducing uniform gain/loss to the left/right portions of the system in Eq.(3), i.e. $V_n \to V_n - i\gamma$ and $\bar{V}_n \to \bar{V}_n + i\gamma$. Because of the simplicity in its structure, this model allows reaching higher generations for more accurate numerical analyses compared to the computationally costly finite element models in previous three examples.

Consider the parametric evolution of frequencies of the P-symmetric model as the coupling constant $t$ that connects the two Fibonacci sub-systems increases. For $t$=0, we have two replicas of the same Fibonacci chain in Eq.(2) and, therefore, the spectrum consists of pairs of degenerate modes. As the coupling $t$ increases, the degeneracy is lifted $\omega_n^{\pm} = \omega_n \pm t$. Simple degenerate perturbation theory with respect to $t$ indicates that the new eigenstates are a linear symmetric/antisymmetric combination of the eigenstates of the Fibonacci Hamiltonian in Eq.(2). The above perturbative framework is applicable as long as the $t$ is smaller than the distance between nearby frequencies $s_n = \omega_{n+1} - \omega_n$ of the uncoupled Hamiltonian $H$ in Eq.(2). The frequency clustering occurring for fractal spectra, however, enforces a rapid breakdown of the perturbation theory, even for infinitesimal $t$. Nevertheless, the eigenstates of the Hamiltonian $H_P(t)$ are still eigenfunctions of the P-symmetric operator and therefore are symmetric or anti-symmetric with respect to the mirror axis of the total chain. The frequency spacing of nearby levels, however, is not dictated by $t$ but the fractal nature of the spectrum.

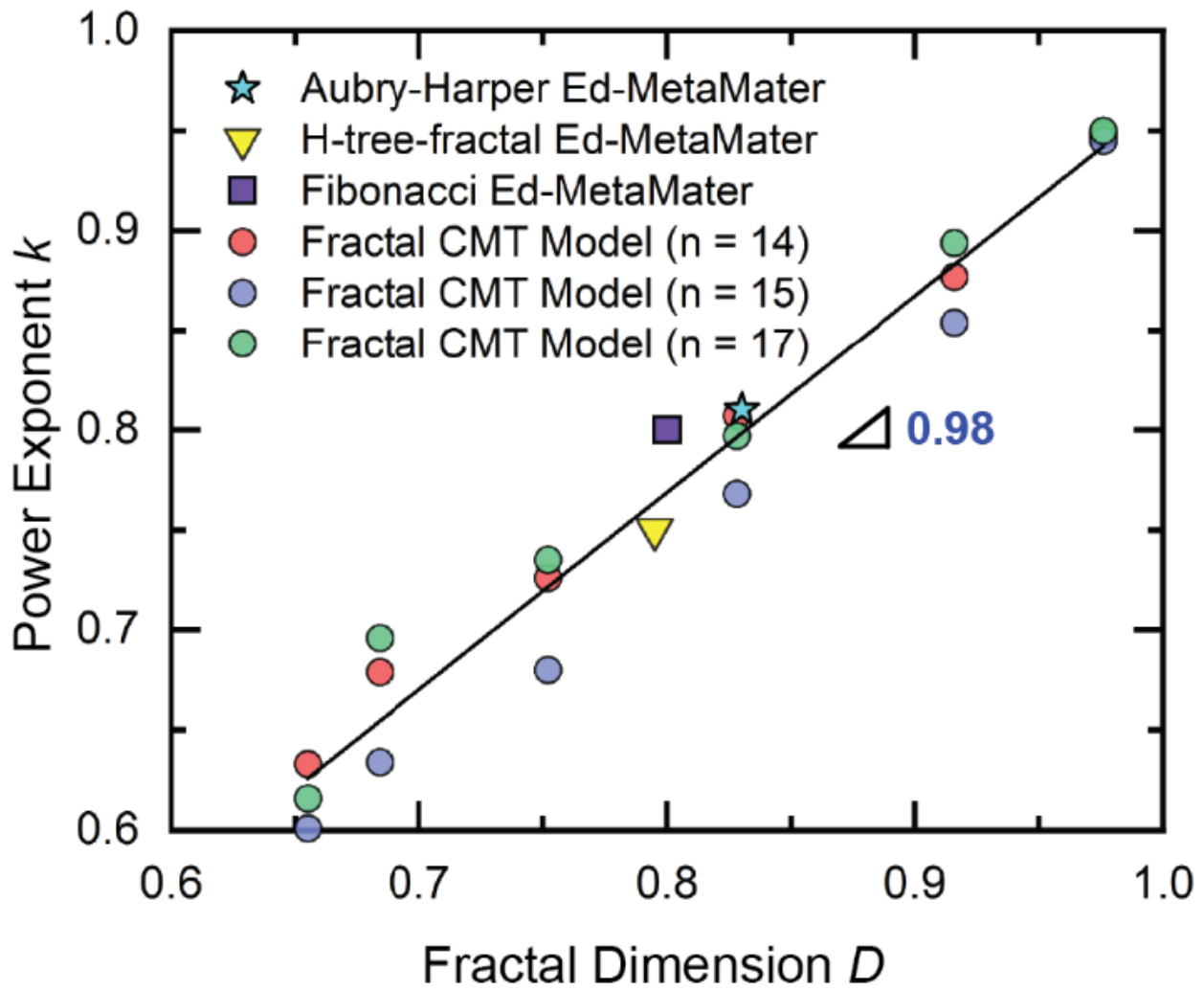

FIG.4. The universal relations between the best-fit power exponents $k$ of the integrated distribution $\mathcal{N}(\Delta_0) \sim 1/\Delta_0^k$ and the spectral fractal dimensions $D$ of various fractal metamaterials.

We treat the inclusion of a small non-Hermitian element $\pm\gamma$ perturbatively. In this case the total Hamiltonian $H_{PT}$ can be written as $H_{PT} = H_P(t) + i\gamma\Gamma$ where the $2N \times 2N$ perturbation matrix $\Gamma$ has elements $\Gamma_{nm} = \delta_{nm}$ for $n \leq -1$ and $\Gamma_{nm} = -\delta_{nm}$ for $n \geq 1$. Finite $\gamma$ leads to level shifts proportional to $\gamma^2$ since the first-order correction vanishes due to the P-symmetry of the corresponding unperturbed eigenmodes of $H_P(t)$. For $\gamma = \gamma_{EP} \simeq s = \Delta_0$, the perturbation theory breaks down, signaling level crossing and the appearance of pairs of complex frequencies. We tested the linear relation $\gamma_{EP} \sim \Delta_0$ for a variety of $V$-values and find that the linear relation holds with a good approximation in all cases (Fig.3(a)). In case of finite system sizes $N$, some frequency differences $\Delta_0$ are still dictated by $t$, though their weight goes to zero at the thermodynamic limit $N \to \infty$. The above analysis allows us to associate the PDF of the gain/loss intensity that results in EP degeneracy with the distribution of level spacings, leading to the conclusion that $\mathcal{P}(\gamma_{EP}) \sim \gamma_{EP}^{-(1+D)}$. We tested the validity of the above arguments numerically using the Fibonacci CMT model for a variety of potentials $V$ and corresponding fractal dimensions $D(V)$ and in all cases we find an excellent agreement with the above theoretical results (Fig.3(b)).

The Fig.4 comprehensively presents the relationship between the spectral fractal dimensions of all aforementioned P-symmetric systems and the power exponents corresponding to the EPs in the PT-symmetric Fibonacci CMT model with different on-site resonances (indicated by circles; further details in Fig.S7 and Fig.S8 [45]), the PT-symmetric Aubry-Harper Ed-MetaMater (blue star), the PT-symmetric H-tree-fractal Ed-MetaMater (yellow triangle), and the PT-symmetric Fibonacci Ed-MetaMater (purple square). The universality in the relations between the emergence of EPs in these metamaterials and the fractality of their initial spectra is evident in Fig.4. The linear fit (black line) with a slope ~1 signifies the universality of this equality relationship, i.e. the power-law exponent describing a scale-free PDF $\mathcal{P}(\gamma_{EP}) \sim \gamma_{EP}^{-(1+D)}$ in a PT-symmetric Ed-MetaMater with unfolding spectral symmetries can directly be obtained from its spectral fractal dimension $D$. This enables a universal route for effectively predicting the emergence of EPs by the initial spectrum itself.

## Conclusions

In summary, we designed three PT-symmetric metamaterials with fractal frequency spectrum—a quasi-periodic Aubry-Harper Ed-MetaMater, an H-tree geometric fractal Ed-MetaMater, and an aperiodic Fibonacci Ed-MetaMater—and investigated them using steady-state dynamic finite element approach. The scale-free emergence of numerous EPs is seen in all metamaterials, showing an intimate relation between the scale-free distribution of critical gain/loss intensities and the spectral fractal dimension of the corresponding Hermitian spectra. Particularly, the linear relation found between the critical gain/loss required for creating EPs and the initial split between the mode pairs that coalesce, shows that the high-signal-quality hypersensitive sensors that exploit EPs in PT-symmetric metamaterials can be engineered by appropriate interacting mode pairs that facilitate experimentally realizable low gain/loss. We further verified those findings from the specific classes of quasi-periodic, fractal, and aperiodic metamaterials and generalized them to a universal law using a CMT-based PT-symmetric fractal mathematical model. The universal relations among the creation of EPs, the scale-free probability distribution of critical gain/loss intensity, and the fractal dimension of the underlying Hermitian spectrum in these PT-symmetric Ed-MetaMater provide a powerful and convenient tool for predicting the emergence of EPs. Our findings are applicable beyond the elastodynamic realm to PT-symmetric metamaterials in acoustic, optical, microwave, and radiofrequency domains as well.

**Supplemental Material**

**for**

# Universal Route for the Emergence of Exceptional Points in PT-Symmetric Metamaterials with Unfolding Spectral Symmetries

**Yanghao Fang[1], Tsampikos Kottos[2,†], and Ramathasan Thevamaran[1,3,*]**

[1]*Department of Materials Science and Engineering, University of Wisconsin-Madison, Madison, Wisconsin 53706, USA*
[2]*Wave Transport in Complex Systems Lab, Department of Physics, Wesleyan University, Middletown, Connecticut 06459, USA*
[3]*Department of Engineering Physics, University of Wisconsin-Madison, Madison, Wisconsin 53706, USA*

Emails: [†]tkottos@wesleyan.edu; [*]thevamaran@wisc.edu

## Finite Element Modeling

We used a commercial finite element platform (*Abaqus Simulia*) to computationally model the steady-state dynamics of the Ed-MetaMater. The material properties of aluminum (Young's modulus: 68.9 GPa, Poisson's ratio: 0.33, density: 2700 kg/m$^3$) is assumed for all components of the Ed-MetaMater which are modeled with cylindrical beam elements (B32). The P-symmetric Ed-MetaMater is harmonically excited using a prescribed axial displacement at the left end of the horizontal coupling rod and the corresponding sinusoidal axial reaction force at its fixed-right-end is measured, simulating a steady-state elastic wave propagation in the metamaterial. The PT-symmetry is created by introducing at each P-symmetric part of the Ed-MetaMater equal amount of energy amplification and attenuation, characterized by an amplification/attenuation rate. These gain/loss mechanisms have been modeled by introducing a structural anti-damping/damping coefficient in *Abaqus* with its magnitude varying from $\gamma = 0.001$ to 1.

## Fibonacci Substitutional Rule and Coupled-mode-theory Modeling for PT-symmetric Fractal Metamaterials

Based on Fibonacci sequence, the number of on-site potentials in the Gen-(n+2) follows:

$$N_{n+2} = N_{n+1} + N_n$$

with $N_0 = 0$ and $N_1 = 1$. In this mathematical model, we use '$A$' and '$B$' to represent two different on-site potentials $-V$ and $+V$ that form the Hamiltonian (correspond to solid and hollow circles in Fig.1(d)).

The substitutional rule for Fibonacci sequence is:

$$A \rightarrow B$$
$$B \rightarrow BA$$

and the first five generations with the starting generation containing only $A$ can be formulated as follows:

| | Chain Sequence | Generation |
|---|---|---|
| $n = 1$ | $A$ | Gen-1 |
| $n = 2$ | $B$ | Gen-2 |
| $n = 3$ | $BA$ | Gen-3 = Gen-2⊕Gen-1 |
| $n = 4$ | $BAB$ | Gen-4 = Gen-3⊕Gen-2 |
| $n = 5$ | $BABBA$ | Gen-5 = Gen-4⊕Gen-3 |

Where ⊕ indicates an operator which links two sequences (for example $xy \oplus abc = xyabc$).

In order to make the model PT-symmetric, a 'mirrored' chain is coupled with its original counterpart and $B = V \pm i\gamma$ and $A = -V \pm i\gamma$ are assumed where $V \in [0, 1]$ is the on-site potential and the imaginary part $\gamma$ is the gain/loss intensity. The sign of $i\gamma$ is negative for $A$ and $B$ that are in the upper half the matrix (represents the gain component), and positive for $A$ and $B$ that are in the lower half the matrix (represents the loss component).

The first five generations of such PT-symmetric chain are shown below:

| | Chain Sequence | Mirrored Chain Sequence |
|---|---|---|
| Substitutional Rule | $A \rightarrow B, B \rightarrow BA$ | $A \rightarrow B, B \rightarrow AB$ |
| $n = 1$ | $A$ | $A$ |
| $n = 2$ | $B$ | $B$ |
| $n = 3$ | $BA$ | $AB$ |
| $n = 4$ | $BAB$ | $BAB$ |
| $n = 5$ | $BABBA$ | $ABBAB$ |

Fig.1(d) shows an illustration of this coupled model. In matrix notation, these two mirrored chains are orderly located along the diagonal with 1 as the neighboring element on either side and all other elements of the matrix are 0 (here, 1 represents the nearest neighbor coupling). For example, the Hamiltonian of the generations $n = 3, 4, 5$ PT-symmetric fractal models are shown below:

$$H_{n=3} = \begin{pmatrix} B & 1 & 0 & 0 \\ 1 & A & 1 & 0 \\ 0 & 1 & A & 1 \\ 0 & 0 & 1 & B \end{pmatrix}$$

$$H_{n=4} = \begin{pmatrix} B & 1 & 0 & 0 & 0 & 0 \\ 1 & A & 1 & 0 & 0 & 0 \\ 0 & 1 & B & 1 & 0 & 0 \\ 0 & 0 & 1 & B & 1 & 0 \\ 0 & 0 & 0 & 1 & A & 1 \\ 0 & 0 & 0 & 0 & 1 & B \end{pmatrix}$$

$$H_{n=5} = \begin{pmatrix} B & 1 & 0 & 0 & 0 & 0 & 0 & 0 & 0 & 0 \\ 1 & A & 1 & 0 & 0 & 0 & 0 & 0 & 0 & 0 \\ 0 & 1 & B & 1 & 0 & 0 & 0 & 0 & 0 & 0 \\ 0 & 0 & 1 & B & 1 & 0 & 0 & 0 & 0 & 0 \\ 0 & 0 & 0 & 1 & A & 1 & 0 & 0 & 0 & 0 \\ 0 & 0 & 0 & 0 & 1 & A & 1 & 0 & 0 & 0 \\ 0 & 0 & 0 & 0 & 0 & 1 & B & 1 & 0 & 0 \\ 0 & 0 & 0 & 0 & 0 & 0 & 1 & B & 1 & 0 \\ 0 & 0 & 0 & 0 & 0 & 0 & 0 & 1 & A & 1 \\ 0 & 0 & 0 & 0 & 0 & 0 & 0 & 0 & 1 & B \end{pmatrix}$$

## Supplementary Figures

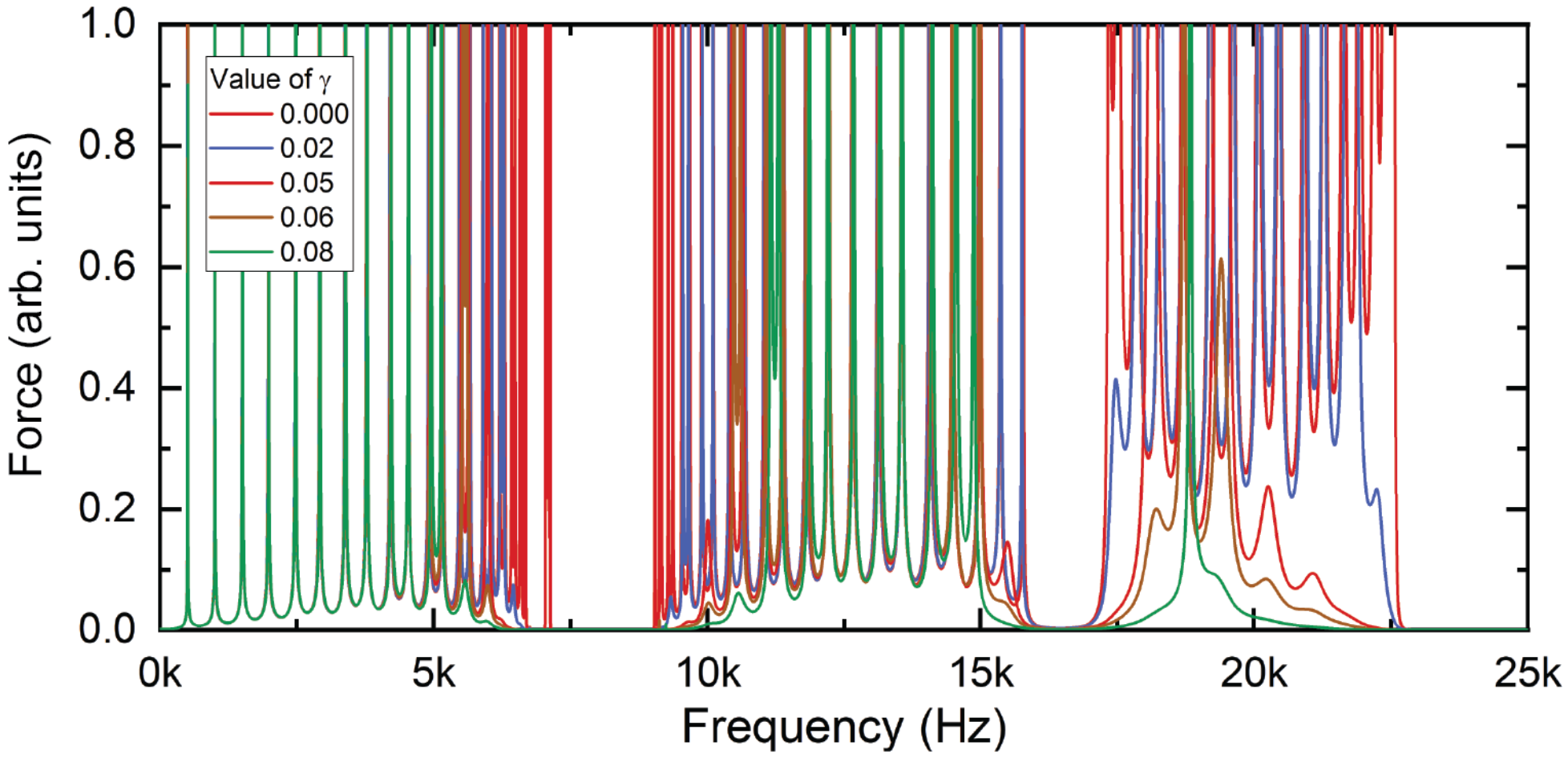

FIG. S1. Magnitude of axial reaction force from a Gen-10 PT-symmetric Aubry-Harper Ed-MetaMater with five different representative values of gain/loss intensity $\gamma$. The mode shifting, coalescing, and eventual damping can be observed with the increasing $\gamma$.

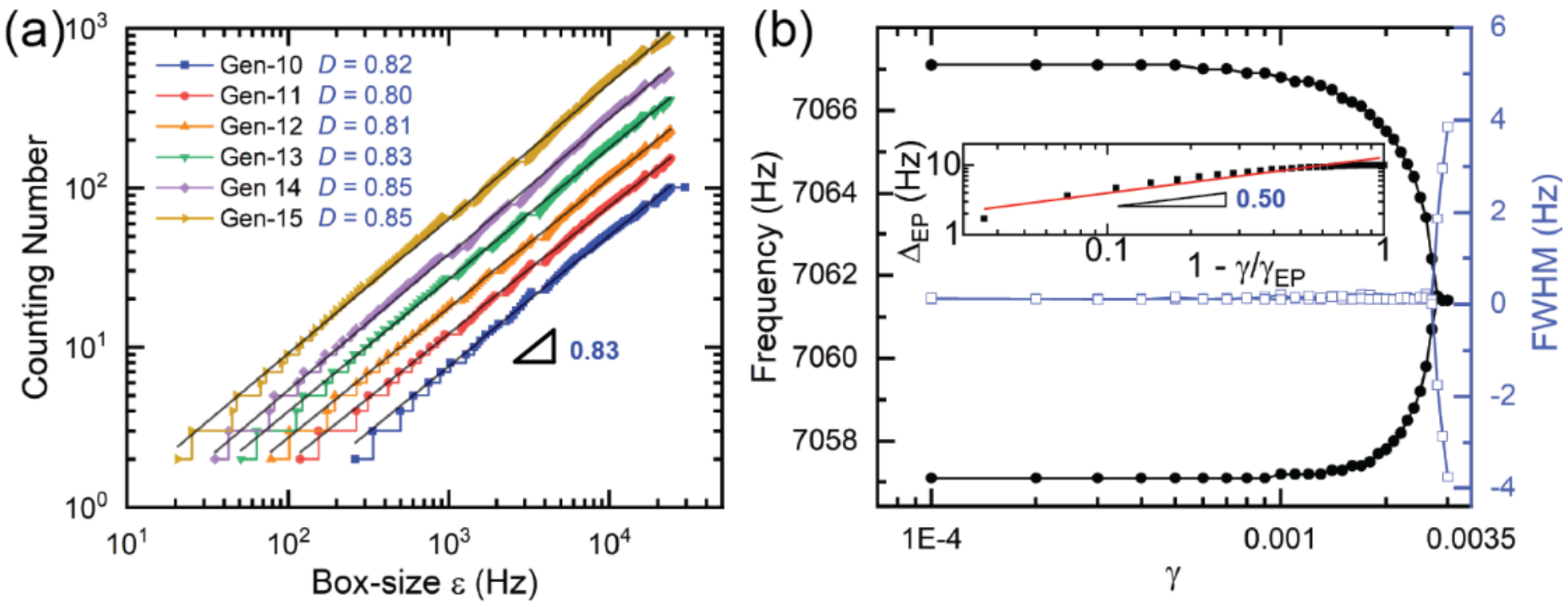

FIG. S2. (a) The spectral fractal dimensions for various generations in PT-symmetric Aubry-Harper Ed-MetaMater (average ~0.83). (b) A characteristic EP formation in a Gen-10 Ed-MetaMater (black and blue symbols correspond to the real and the imaginary parts of the modes, respectively); inset shows the $\Delta_{EP}$ vs. $1 - \gamma/\gamma_{EP}$.

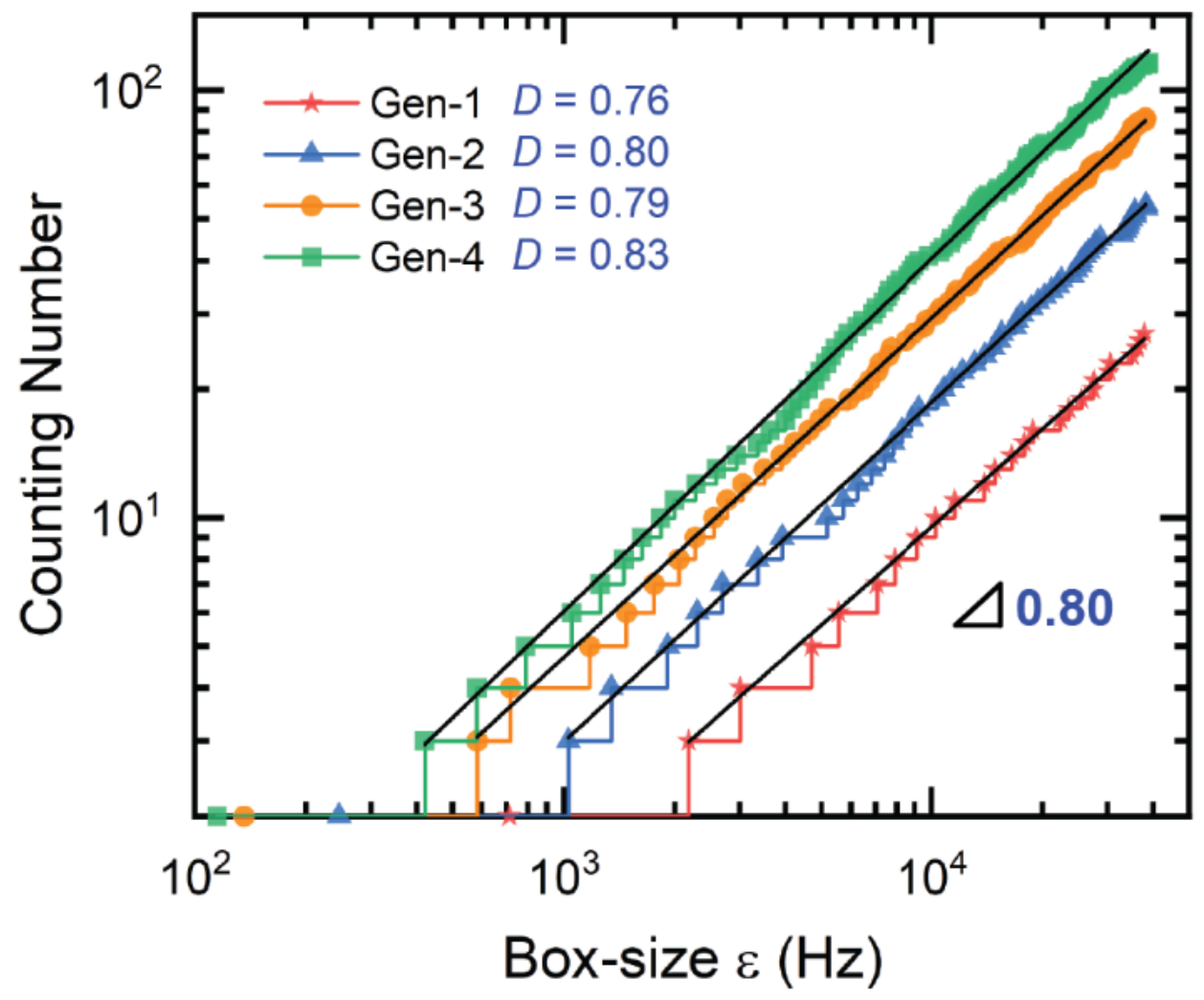

FIG. S3. The spectral fractal dimensions for the first four generations of PT-symmetric H-tree fractal Ed-MetaMater (average ~0.80).

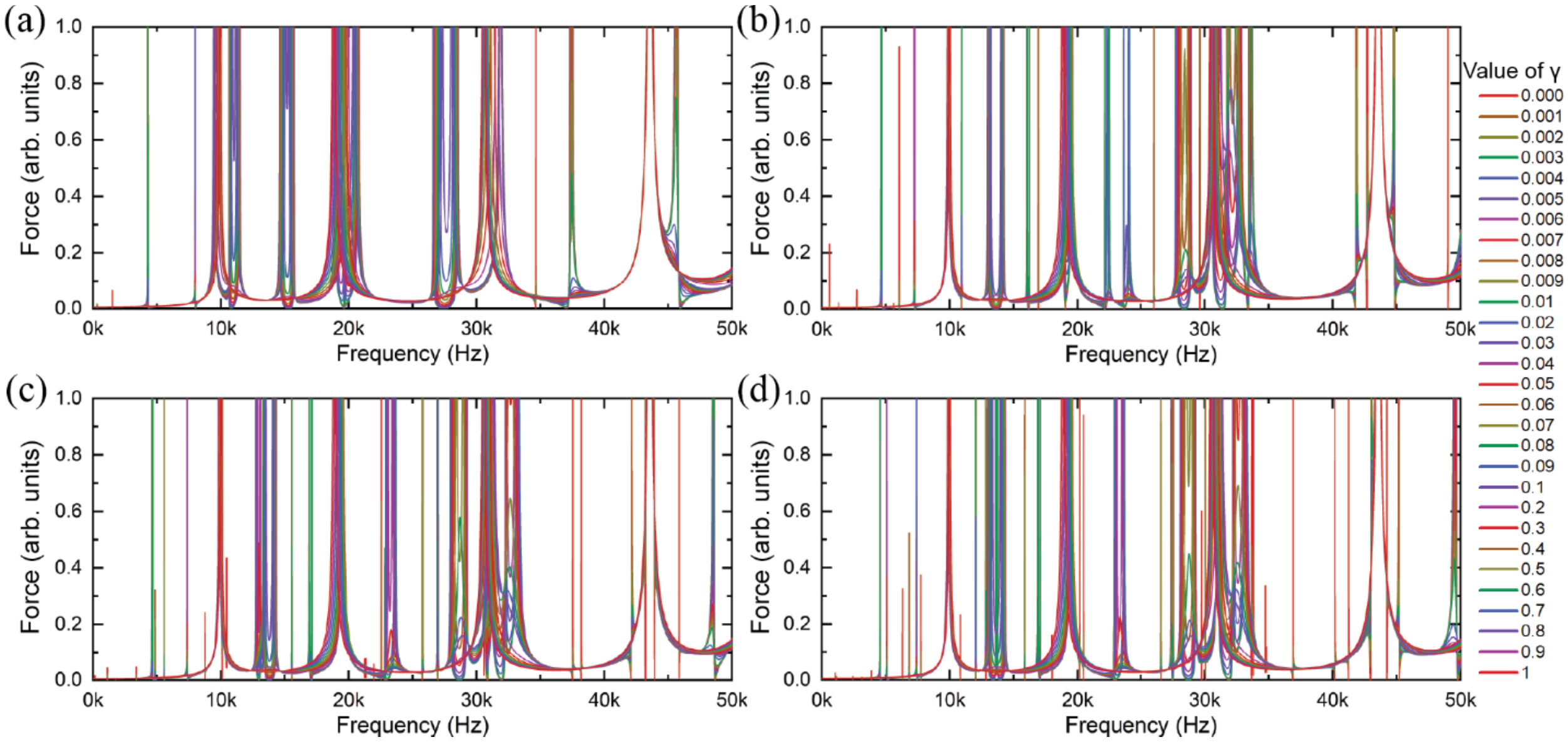

FIG. S4. Magnitude of axial reaction force from (a) Gen-1, (b) Gen-2, (c) Gen-3, and (d) Gen-4 quasi-3D PT-symmetric H-tree fractal Ed-MetaMater with gain/loss intensity $\gamma$ varied from 0 to 1.

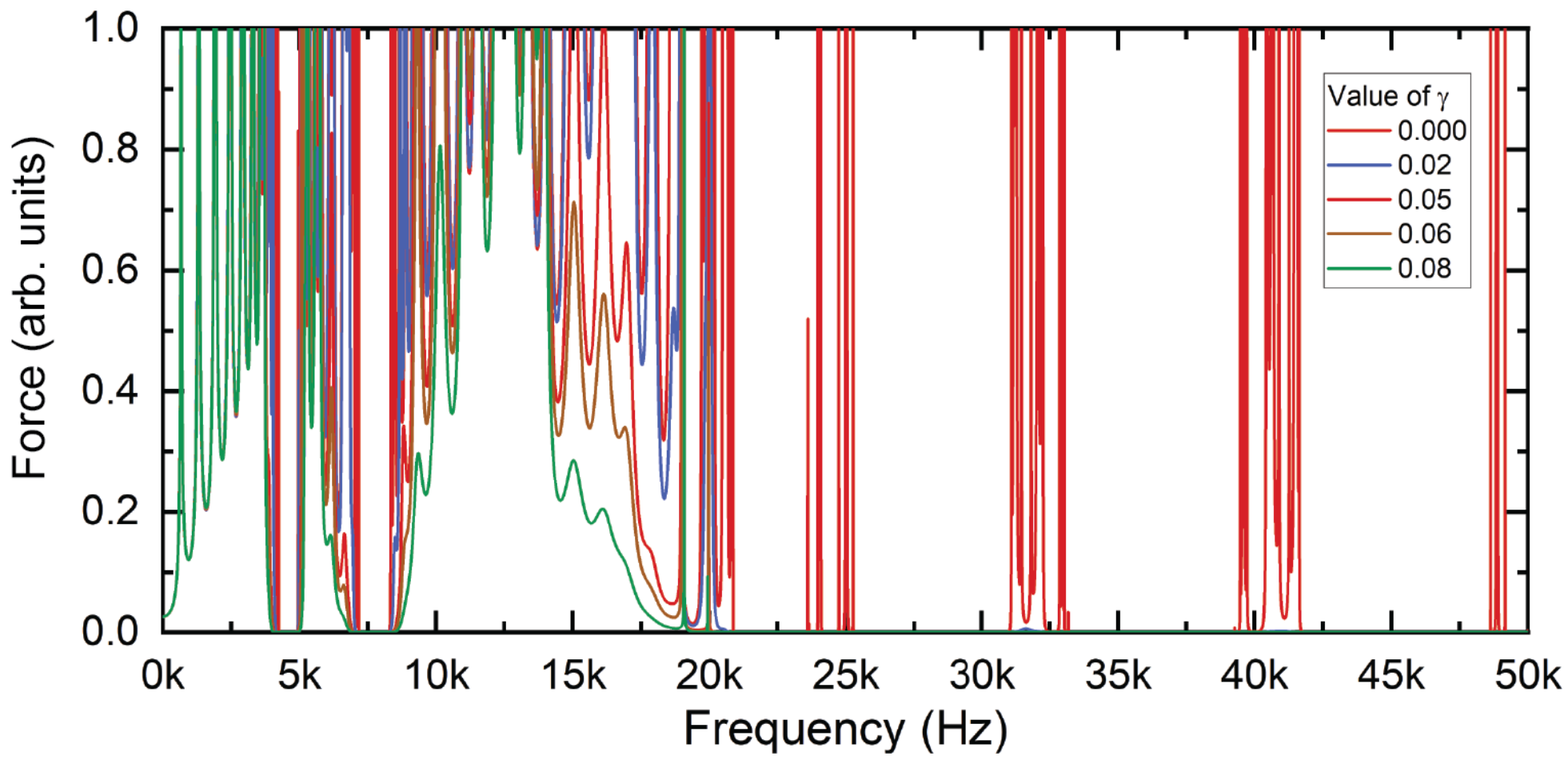

FIG. S5. Magnitude of axial reaction force from a Gen-9 PT-symmetric Fibonacci Ed-MetaMater with five different representative values of gain/loss intensity $\gamma$.

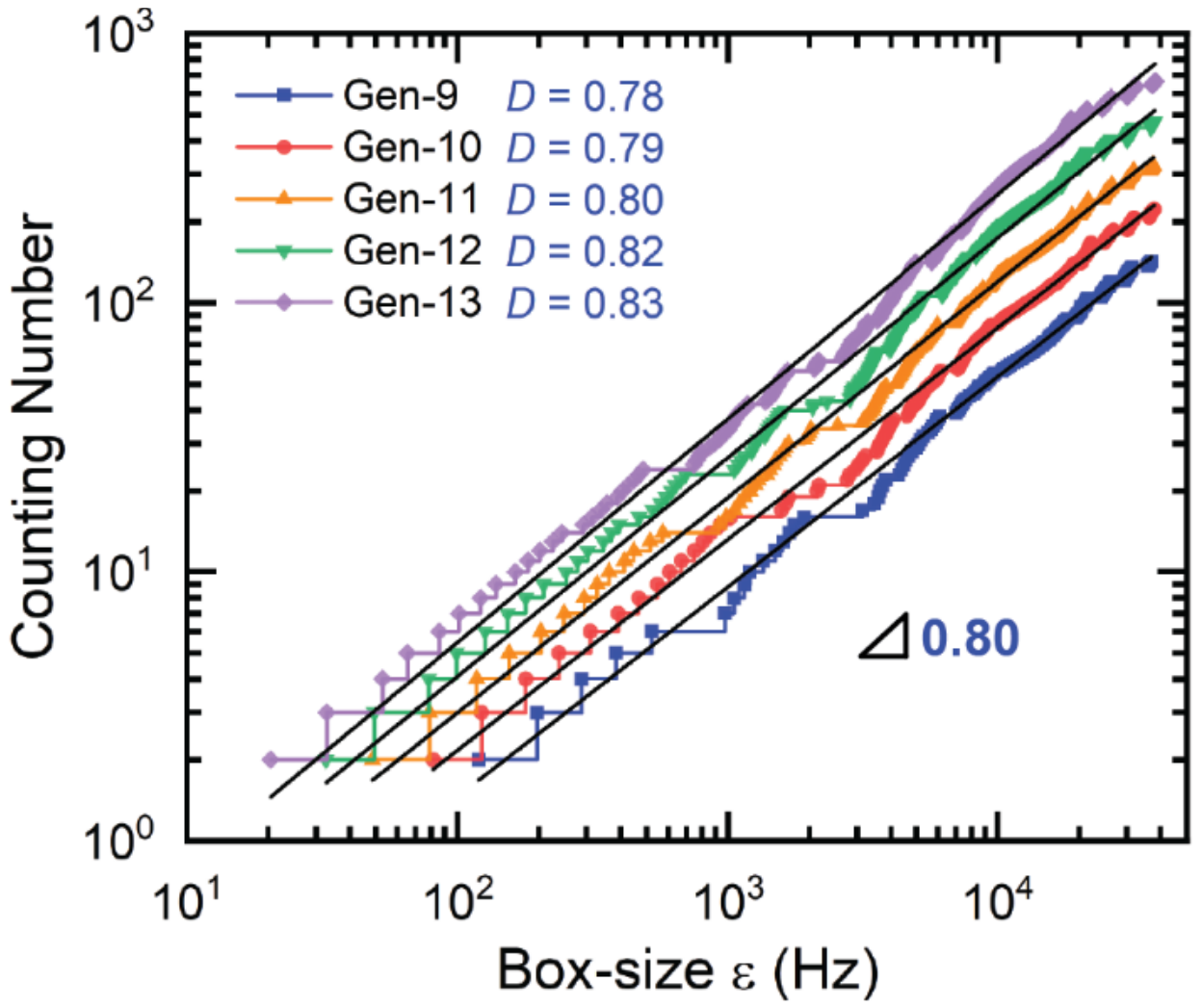

FIG. S6. The spectral fractal dimensions for various generations of PT-symmetric Fibonacci Ed-MetaMater (average ~0.80).

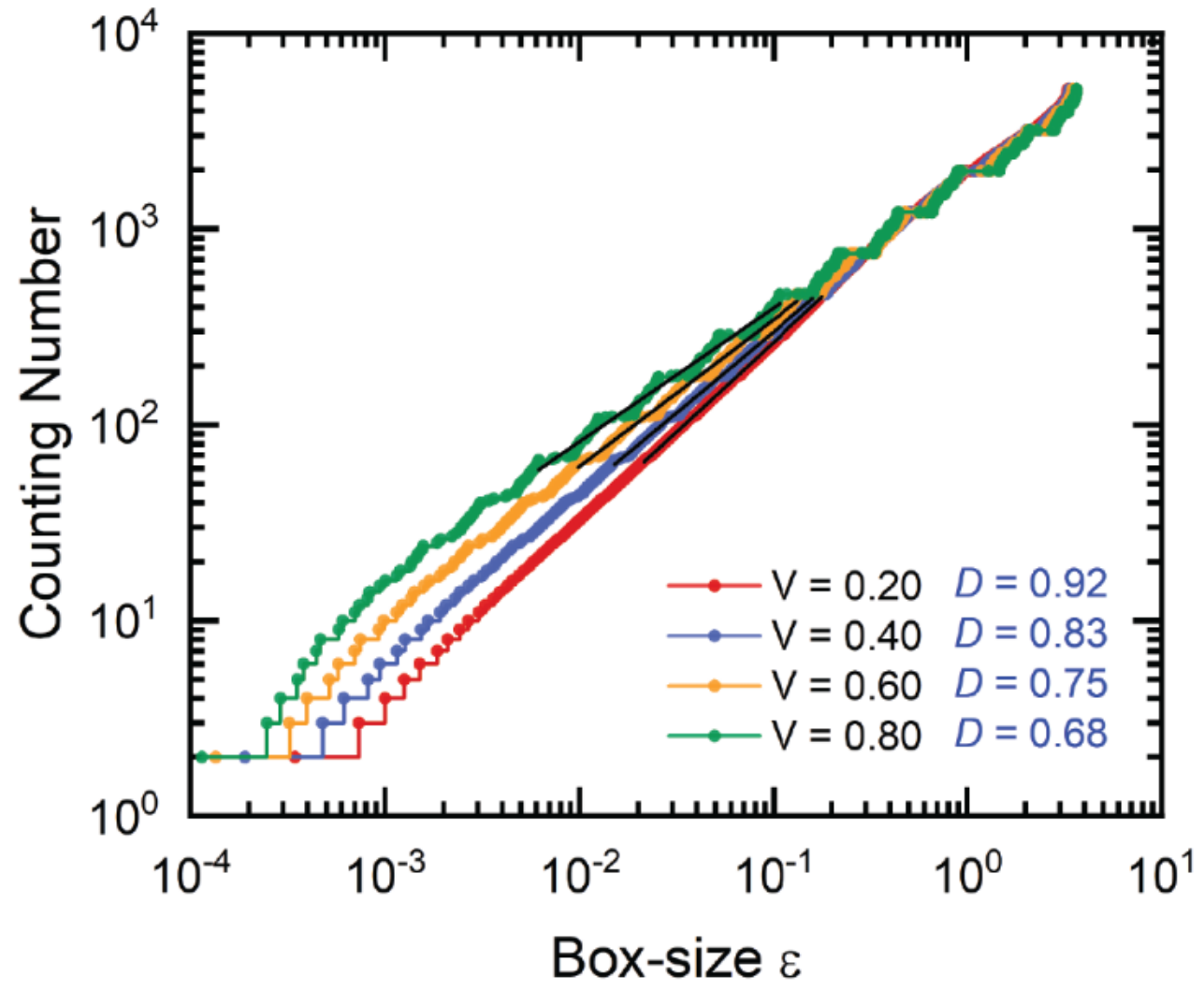

FIG. S7. The spectral fractal dimensions of different CMT-based mathematical models. They varied by the magnitude of $V$.

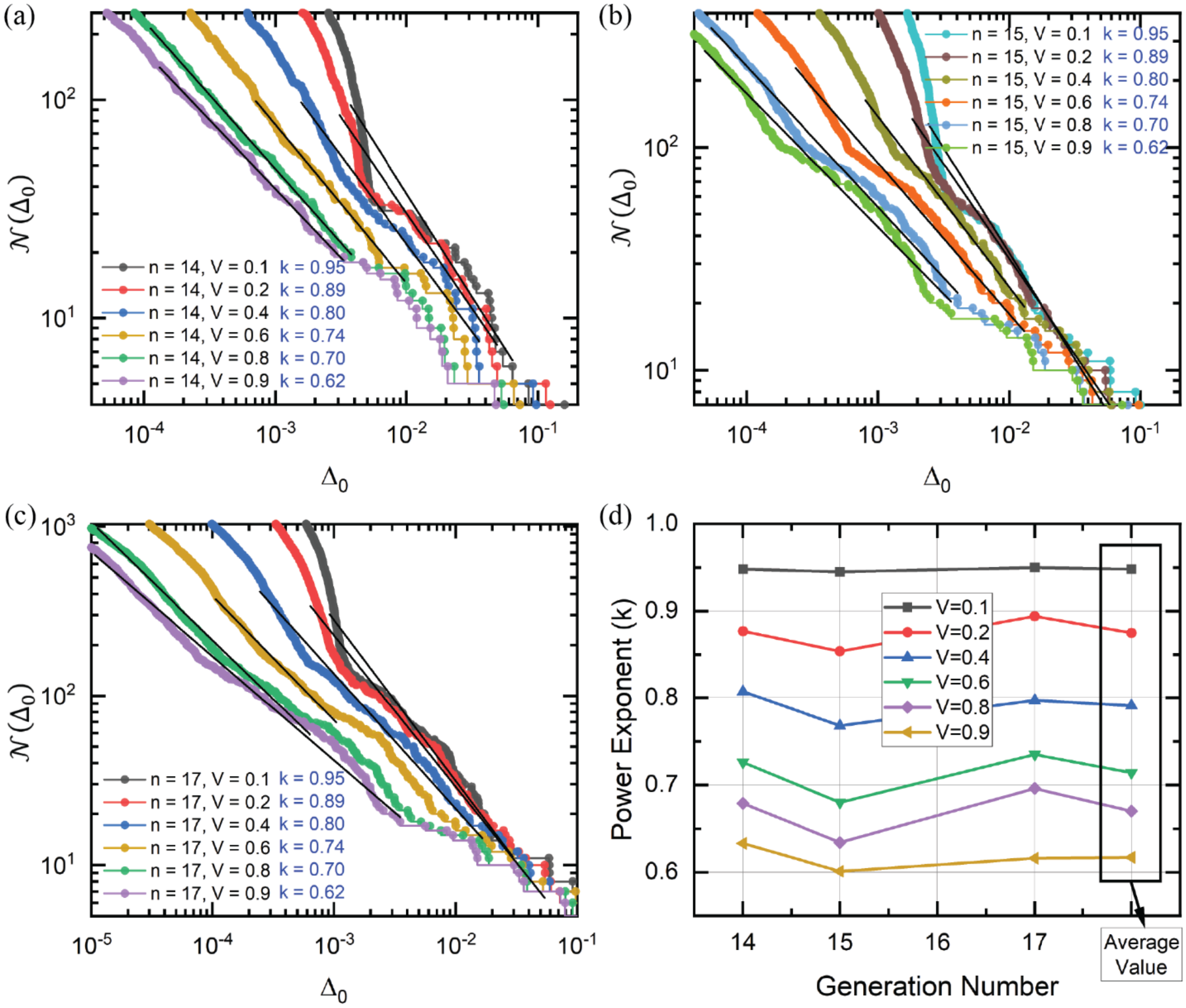

FIG. S8. Integrated distribution $\mathcal{N}(\Delta_0)$ as functions of $\Delta_0$ in (a) Gen-14, (b) Gen-15, and (c) Gen-17 CMT-based mathematical models with six different values of $V$. (d) The power exponents $k$ extracted from the Figs. S8(a-c).